\documentstyle[epsfig]{kamioka}
\begin{document}

\title{%
Neutrino oscillations in high energy cosmic neutrino flux
}

\author{%
Osamu Yasuda \\
{\it Department of Physics, Tokyo Metropolitan University,\\
1-1 Minami-Osawa Hachioji, Tokyo 192-0397, Japan, \\
yasuda@phys.metro-u.ac.jp}}
\maketitle

\section*{Abstract}

I discuss the effects of neutrino oscillations on high energy cosmic
neutrinos which come from cosmologically distant astrophysical
sources.  I incorporate all the up-to-date constraints from the solar,
atmospheric, reactor, accelerator data and give the possible pattern
for the ratio of the high energy cosmic neutrinos in the cases of
three and four neutrino schemes.

\section{Introduction}

A lot of attention has been focused on high energy neutrinos ($E\,
\geq 10^{6}$ GeV) which come from cosmologically distant astrophysical
sources such as Active Galactic Nuclei and Gamma Ray Burst
fireballs (typical distance is 100 Mpc), since they can be
distinguished from atmospheric neutrinos for such high energies and
identification of flavors of neutrinos may be possible in new km$^{2}$
surface area neutrino telescopes \cite{learnedpakvasa,observ}.  The
effects of neutrino oscillations on the high energy cosmic neutrinos
have been discussed in the past \cite{learnedpakvasa,athar,bentoetal},
and the purpose of this talk is to update the analysis by taking into
account all the constraints from the solar, atmospheric, reactor and
accelerator data in the three or four neutrino framework
(This talk is based on the work \cite{atharetal}).

\section{Analysis}
Since the path length of neutrinos is much
larger than any possible neutrino oscillation length which is
suggested from the solar, atmospheric or LSND data for the energy
$E_\nu\sim$ 10$^6$ GeV, I will average over rapid oscillations
throughout this talk.  Then the oscillation probability is
given by
\begin{eqnarray}
P(\nu_\alpha\rightarrow\nu_\beta;L=\infty)=\delta_{\alpha\beta}
-\sum_{j\ne k}U_{\alpha j}^\ast U_{\beta j}
U_{\alpha k} U_{\beta k}^\ast
&=&\sum_{j}|U_{\alpha j}|^2 |U_{\beta j}|^2,
\label{eqn:p}
\end{eqnarray}
where $U_{\alpha k}$ stands for the MNS mixing matrix and
I have ignored odd functions in $\sin(\Delta m_{ij}^2L/4E)$
which oscillate rapidly as $L\rightarrow\infty$.

The electron and muon neutrinos are mainly produced in
the decay chain of charged pions whereas the tau neutrinos are mainly
produced in the decay chain of charmed mesons
at a suppressed level \cite{waxman}.
The ratio of the intrinsic
high energy cosmic neutrinos flux is typically
 $F^{0}(\nu_{e})$ :
$F^{0}(\nu_{\mu})$ : $F^{0}(\nu_{\tau})\, =\, 1$ : $2$ : $<10^{-5}$. For
simplicity I assume that the ratio is
 $F^{0}(\nu_{e})$ : $F^{0}(\nu_{\mu})$ :
 $F^{0}(\nu_{\tau})\, =\, 1$ : $2$ : 0.
Thus the ratio of flux of neutrinos
in the far distance is given by
\begin{eqnarray}
\left( \begin{array}{c} F(\nu_e)  \\ F(\nu_\mu) \\ 
F(\nu_\tau) \end{array} \right)
=P\left( \begin{array}{c} F^0(\nu_e)  \\ F^0(\nu_\mu) \\ 
F^0(\nu_\tau) \end{array} \right)
=P\left( \begin{array}{c} 1  \\ 2 \\ 
0 \end{array} \right)F^0(\nu_e),
\label{eqn:f}
\end{eqnarray}
where a matrix $P$ has components
$\left(P\right)_{\alpha\beta}=P(\nu_\alpha\rightarrow\nu_\beta;L=\infty)$
(See (\ref{eqn:p})).

Since currently we do not know the precise total cosmic neutrino flux
I will mainly focus my discussion on the ratio of different flavors
of neutrinos.
To plot the ratio of the three neutrino flavors, 
I introduce a triangle
representation.  Fig. 1 is a unit regular triangle
and the position of the point gives the ratio of the high energy 
neutrino flux, where
$F_\alpha\equiv F(\nu_\alpha)$ is given by (\ref{eqn:f}).

\vglue 0.5cm
\begin{figure}[h]
\vglue -0.5cm\hglue -0.3cm\epsfig{file=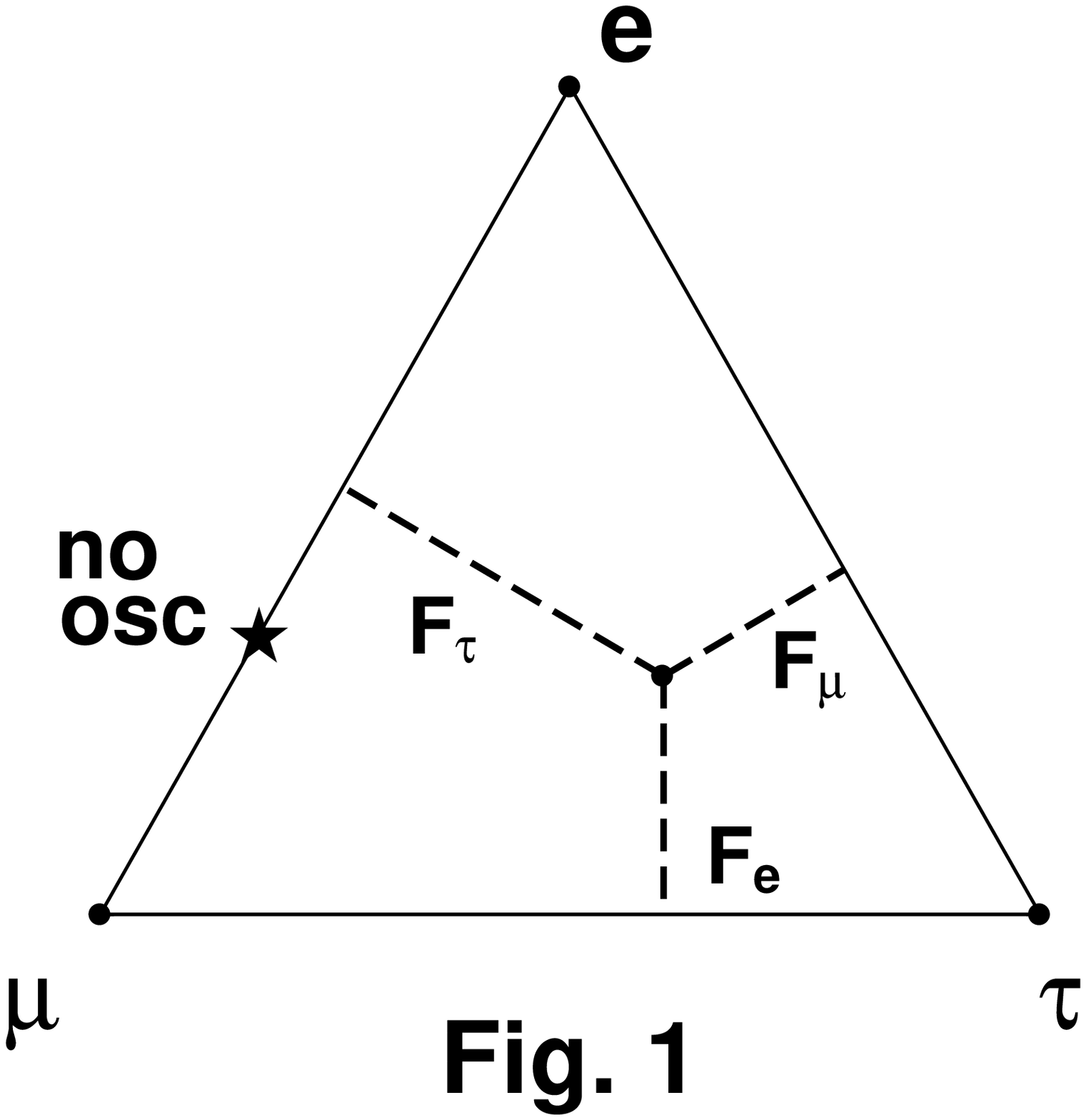,width=0.50\textwidth}
\vglue -7.4cm\hglue 7.5cm
\epsfig{file=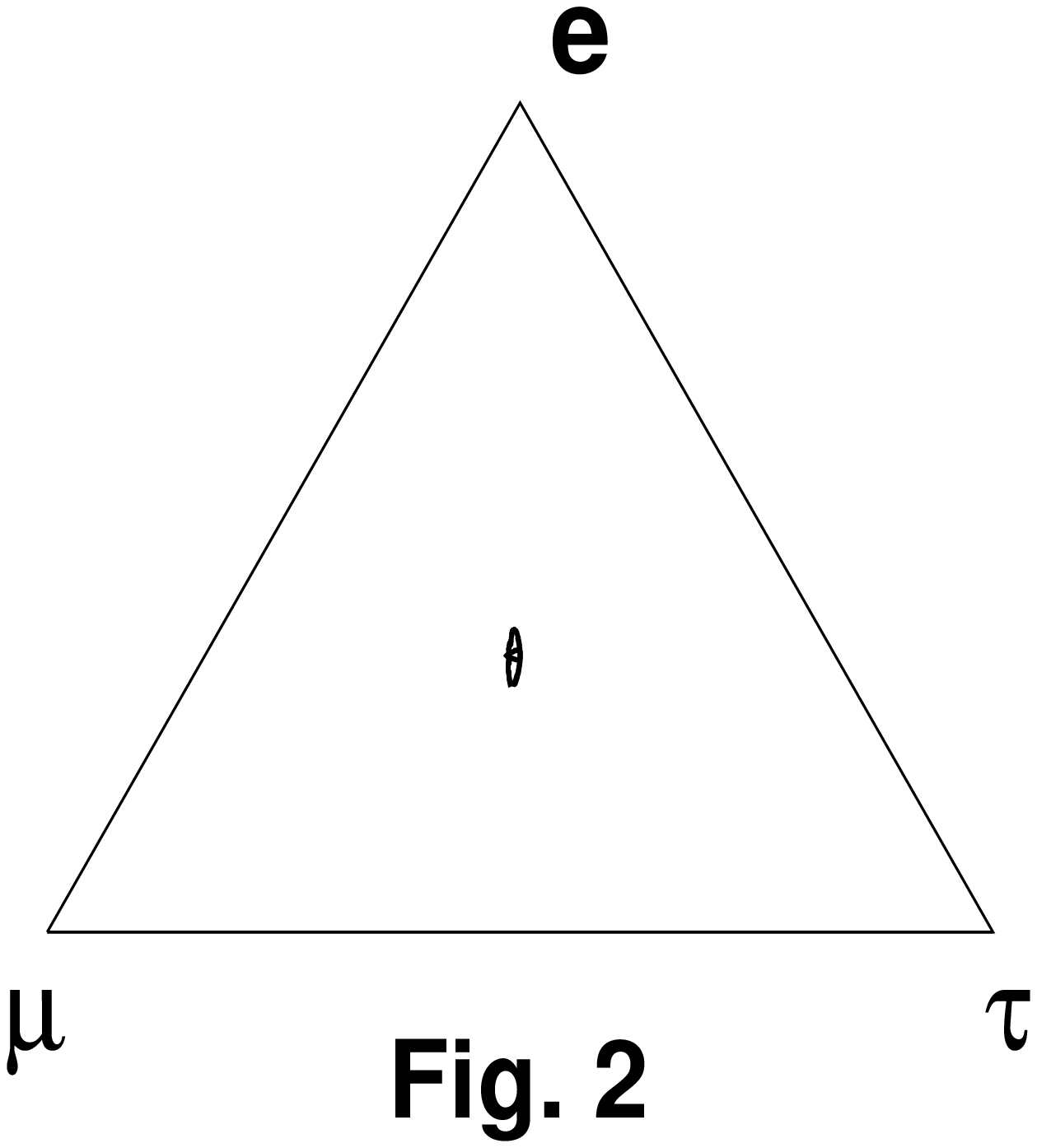,width=0.60\textwidth}
\end{figure}

In the three flavor framework, because of the constraint of
the CHOOZ data \cite{chooz} and the atmospheric neutrino data
of Superkamiokande and Kamiokande, it has been known that
$|U_{e3}|^2$ is small and $|U_{\mu j}|^2 \simeq|U_{\tau j}|^2|$
(See, e.g., \cite {yasuda1}).  Using the allowed region for
$|U_{\alpha j}|^2$ ($\alpha=e,\mu,\tau$) in the atmospheric
neutrino analysis of \cite{yasuda1} and
in solar neutrino analysis of \cite{foglietal}, the possible
ratio of the high energy neutrino flux is calculated numerically and
is given in Fig. 2.  The allowed
region is a small area around the midpoint $F(\nu_e)=F(\nu_\mu)=
F(\nu_\tau)=1/3$.

\begin{figure}[h]
\hglue -0.6cm\epsfig{file=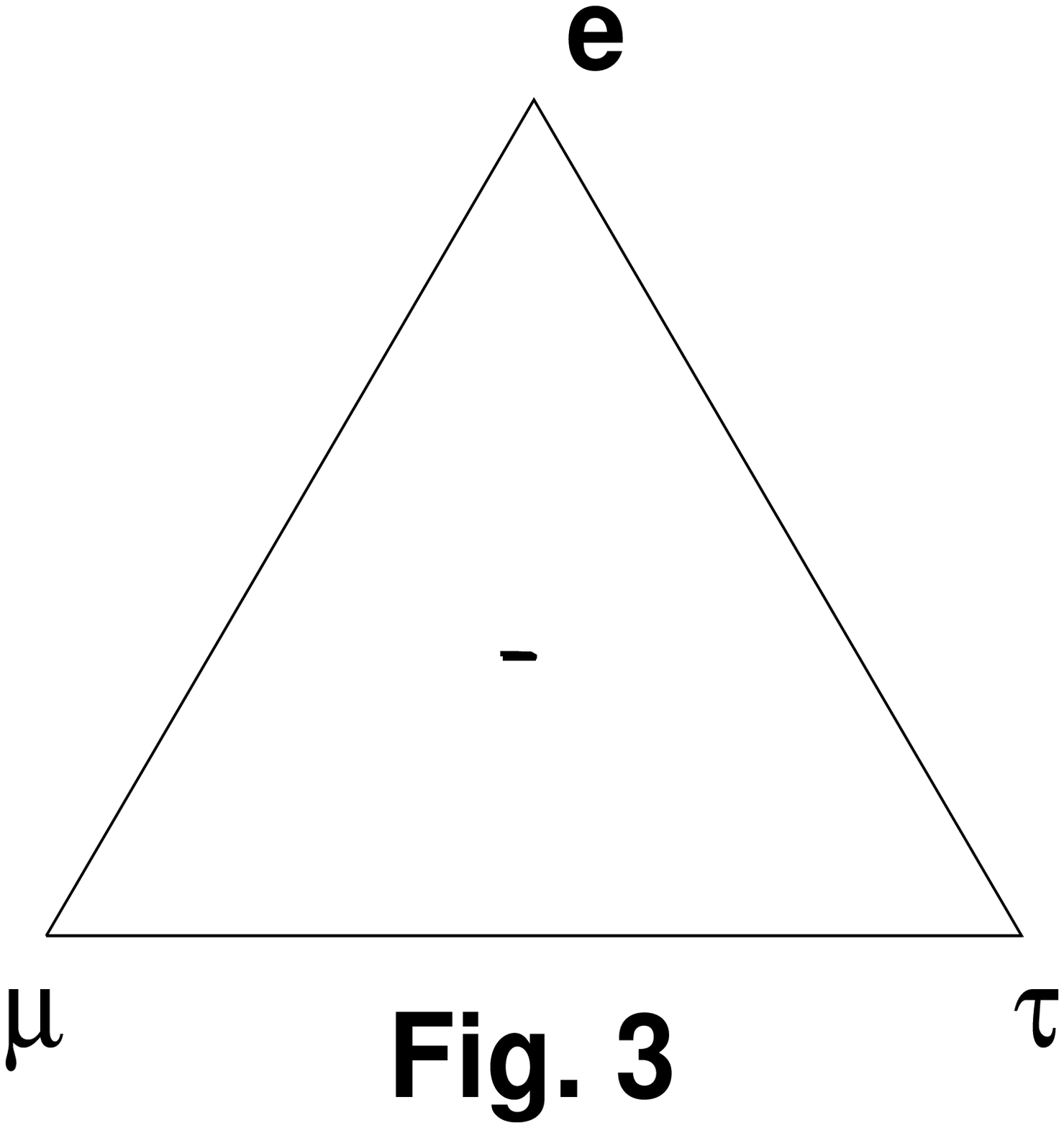,width=0.50\textwidth}
\vglue -7.2cm\hglue 7.2cm
\epsfig{file=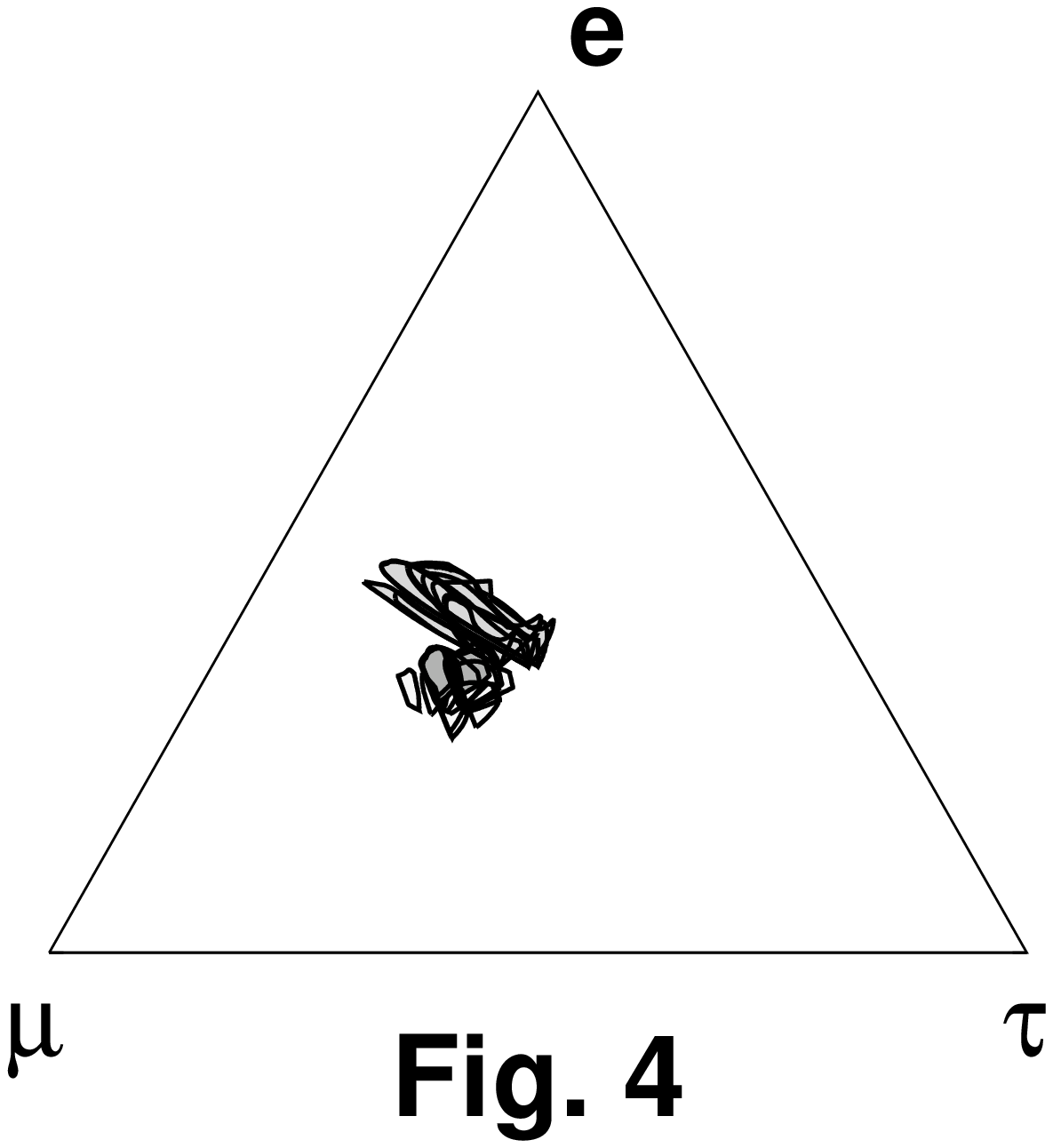,width=0.60\textwidth}
\end{figure}

In the four neutrino scheme one needs in principle tetrahedron to
express the ratio of the four neutrino flux, but since we do not
observe the cosmic sterile neutrino I normalize the flux of
each active neutrino by the total flux of active ones:
\begin{eqnarray} \left( \begin{array}{c} \widetilde{F}(\nu_e) \\
\widetilde{F}(\nu_\mu) \\ \widetilde{F}(\nu_\tau) \end{array} \right)
\equiv{1 \over F(\nu_e)+F(\nu_\mu)+F(\nu_\tau)} \left(
\begin{array}{c} F(\nu_e) \\ F(\nu_\mu) \\ F(\nu_\tau) \end{array}
\right).  \end{eqnarray} After redefining the flux this way
($\widetilde{F}\rightarrow F$), we can plot the ratio of each active
neutrino with the same triangle graph as in the three neutrino case.

If one demands that the number $N_\nu$ of effective neutrinos in Big Bang
Nucleosynthesis (BBN) be less than 4, then it can be shown
\cite{okadayasuda} that
the $4\times 4$ MNS mixing
matrix splits into two $2\times 2$ block diagonal matrices.  In this case
the ratio is given by a small region depicted in Fig. 3.

On the other hand, some people \cite{he4} give conservative bound for $N_\nu$,
and without the BBN constraint $N_\nu < 4$ the only restrictions come from
the solar and atmospheric neutrino data.
The analysis of the solar neutrino data in the four neutrino scheme
with ansatz $U_{e3}=U_{e4}=0$ has been done recently in \cite{giuntietal}.
The analysis of the atmospheric neutrino data in the four neutrino
framework has been done in \cite{yasuda2}
again with ansatz $U_{e3}=U_{e4}=0$.  Using these results,
the ratio of the high energy cosmic neutrinos is evaluated and is
given in Fig. 4 which has much wider allowed region than any other
case.  This scheme may be distinguished from others
if one has good precision in future
experiments.

I would like to thank hospitality of Summer Institute 99 at Yamanashi,
Japan where this work was started.  This research was supported in
part by a Grant-in-Aid for Scientific Research of the Ministry of
Education, Science and Culture, \#12047222, \#10640280.


\end{document}